\documentclass[prl,superscriptaddress,twocolumn,floatfix,letterpaper,showpacs]{revtex4}
\usepackage{amsmath}
\usepackage{graphicx}
\usepackage{times}
\usepackage{hyperref}
\usepackage{subfigure}
\begin{document}

\title{Superionic to superionic phase change in water: consequences for the interiors of Uranus and Neptune }
\author{Hugh F. Wilson, Michael L. Wong} 
\affiliation{Department of Earth and Planetary Science, University of California, Berkeley}
\author{Burkhard Militzer} 
\affiliation{Department of Earth and Planetary Science and Department of Astronomy, University of California, Berkeley}
\begin{abstract}
Using density functional molecular dynamics free energy calculations, we show that the body-centered-cubic (bcc) phase of superionic ice previously believed to be the only phase is in fact thermodynamically unstable compared to a novel phase with oxygen positions in face centered cubic lattice sites. The novel phase has a lower proton mobility than the bcc phase and may exhibit a higher melting temperature. We predict a transition between the two phases at a pressure of 1 $\pm$ 0.5 Mbar, with potential consequences for the interiors of ice giants such as Uranus and Neptune.
\end{abstract} 
\maketitle

Water is one of the most prevalent substances in the universe and exists in a large number of phases over a vast range of temperature and pressure conditions. In addition to the liquid, gas, plasma and many solid phases \cite{militzer-prl-10,mcmahon-prb-11,wang-nature-11, ji-prb-11,hermann-pnas-12}, water also possesses a superionic phase, in which the oxygen atoms occupy fixed lattice positions as in a solid, while hydrogen atoms migrate through the lattice as in a fluid \cite{french-prb-09,redmer-icarus-11}. The superionic phase is predicted to occupy a large section of the ice phase diagram for pressures in excess of 0.5 Mbar and temperatures of a few thousand Kelvin \cite{cavazzoni,goldman,french-prb-09,redmer-icarus-11}. As this regime corresponds to conditions in the interiors of ice giants such as Uranus and Neptune, which are believed to consist largely of water, it is predicted that these planets consist largely of superionic ice \cite{redmer-icarus-11}, making an understanding of the physical and chemical properties of superionic ice vital for understanding the interior structure and evolution of these planets.

Although superionic ice has been extensively studied in \emph{ab initio} theoretical studies \cite{french-prb-09,redmer-icarus-11,goldman,cavazzoni}, all works up to this point have assumed the superionic phase to maintain a body centered cubic (bcc) structure for the oxygen sublattice, as seen in the solid ice VII and ice X phases \cite{polian-prl-84}. In this paper we predict instead, via density functional theory (DFT) free energy calculations, that the bcc phase is thermodynamically unstable relative to a denser face centered cubic (fcc) phase for pressures in excess of $1.0 \pm 0.5$ Mbar. The fcc phase is found to have a lower hydrogen mobility than the bcc phase. The proposed phase transition may intersect with the Neptunian and Uranuian isentropes, suggesting the possibility of a superionic to superionic phase transition in ice giants.

The existence of superionic ice was initially predicted in via DFT-MD simulations by Cavazzoni \emph{et al} \cite{cavazzoni}, by heating of the ice X and ice VII phases of water at pressures in excess of 0.5 Mbar. The bcc oxygen sublattice of the ice X and VII was found to be maintained upon the transition to the superionic phase. Goldman \cite{goldman} \emph{et al} in 2005 studied bonding and diffusion in superionic water, again assuming the oxygen atoms to retain a bcc sublattice. French \emph{et al.} \cite{french-prb-09,redmer-icarus-11} extensively studied the bcc superionic phase and its transition to the fully fluid or plasma regime in which both hydrogens and oxygens become mobile. In repeated simulations, French \emph{et al.}~cooled water from the fully fluid regime and observed the re-formation of a superionic phase with a bcc oxygen sublattice, however the geometric constraints of the unit cell used, with 54 H$_2$O molecules in a cubic cell, mean that the formation of alternative structures whose sublattices do not conform to these constraints is penalized. Experimentally, superionic ice has been observed in laser-heated diamond anvil cell experiments by Goncharov \emph{et al}.~\cite{goncharov-prl-05} who demonstrated spectroscopically a phase transition believed to correspond to the superionic phase at approximately 0.47 Mbar, however these experiments did not provide structural information. Recently, Sugimura \emph{et al} detected a phase of water ice which they characterise as superionic at pressures from 0.3 to 1 Mbar and temperatures of approximately 850~K; X-ray diffraction shows this phase to have a bcc oxygen sublattice.

Hints of the instability of the bcc oxygen sublattice over at least some portion of the superionic ice regime have been observed in several studies. French \emph{et al} \cite{french-prb-09} noted the existence of a region of the phase diagram at low temperature and moderate pressure in which the bcc oxygen sublattice was unstable within molecular dynamics (MD) -- that is, at which the system readily evolves out of the sublattice in timescales accessible to our MD simulations. In the present authors' study of the solubility properties of superionic ice at giant planet core conditions \cite{wilson-astrop-12} we briefly noted that the bcc phase became unstable at two sets of conditions under consideration (10 Mbar with temperatures of 2000 and 3000~K). Recent work on superionic ammonia \cite{ninet-prl-12} has raised the possibility of the existence of phase changes within the superionic regime of NH$_3$. Furthermore, the bcc sublattice of the ice X phase at zero temperature has been shown to become dynamically unstable at pressures in excess of 4 Mbar by Caracas \cite{caracas}, with higher-pressure zero-temperature ice phases having alternative . These factors motivated a formal study of alternative oxygen sublattices in superionic water.

The first stage of our study was to investigate the short-term stability of the bcc lattice as a function of temperature and density, along with that of the fcc and hexagonal close packed (hcp) lattices. The ability of a system to retain a particular structure over the picosecond timescales associated with a molecular dynamics simulation is a necessary but insufficient condition for a structure to represent the thermodynamic ground state. We restricted our attention to these three simple high-symmetry sublattices on the tentative assumption that thermal vibrations and the disordered migration of the hydrogen atoms mean that the oxygen sublattice is likely to possess a configuration with high degree of symmetry. To investigate the short-term stability of a superionic lattice structure, we first undertake a molecular dynamics simulation in which the oxygen atoms remain constrained to lattice positions while the hydrogen atoms move freely to equilibrate at the desired temperature. The constraint on the oxygen atoms is then released and the dynamics continued, with a newly-generated thermal velocity distribution.  We simulated fcc and bcc ice structures in constant cells at temperatures of 2000--5000~K and at nine densities from 3 gcm$^{-3}$ (approximately 1 Mbar) to 11 gcm$^{-3}$ (approximately 40 Mbar). Simulations in this work used the VASP package \cite{vasp}; further details on calculational parameters and convergence tests are given in online supplementary material.
A distortion in the bcc oxygen sublattice was observed at 2000~K for densities of 6 and 7 gcm$^{-3}$ (corresponding to pressures around 9 and 14 Mbar) and also at 6 gcm$^{-3}$ at a temperature of 4000~K. In a later simulation we also found the bcc structure to become distorted at a pressure of 40 Mbar and temperature of 10000K. The fcc superionic structure remained stable in MD under all conditions studied. 

All attempts to simulate superionic ice with an hcp oxygen sublattice rapidly resulted in the hcp oxygen sublattice becoming distorted. Although the hcp and fcc lattices are very similar, differing only by the stacking of layers, and represent equivalent packings of spheres, we note that there is an important difference in their distribution of interstitial sites. In the fcc superionic structure we find that the hydrogen atoms largely concentrate around the tetrahedral interstitial sites, of which there are two per oxygen atom. Due to the different arrangement of oxygen atoms in the hcp crystal, the equivalent tetrahedral interstitial sites are arranged in very closely spaced pairs, making the simultaneous occupation of all tetrahedral sites disfavored.

Having established the short-term stability of the fcc and bcc phases across most of the superionic ice regime, we must now determine which phase has a lower Gibbs free energy $G = U + PV - TS$. We chose several representative points throughout the superionic regime, ranging from pressures of 1 Mbar close to the onset of superionic behavior up to 40 Mbar corresponding to the approximate pressure of Jupiter's core; we chose points at which both bcc and fcc phases were stable, steering clear of the low-temperature regime around $\rho = 6$ gcm$^{-3}$ ($\approx$ 10 Mbar). 

For the computation of Gibbs free energies we adopt a two-step coupling constant integration (CCI) method as recently applied by several authors \cite{morales-pnas-09,wilson-prl-10,wilson-astrop-12,wilson-prl-12} and identical to that described in more detail for our earlier work on superionic water solubility \cite{wilson-astrop-12}. In this scheme the free energy of the system of interest is computed from the free energy of a simpler system whose free energy is known analytically via a thermodynamic integration in which the simpler system is gradually changed into the system of interest. For the analytic system to resemble the superionic phase we choose a system consisting of noninteracting harmonic oscillators at lattice sites for oxygen atoms, and a noninteracting ideal gas for the hydrogen atoms. The difference in Helmholtz free energy between systems 1 and 2 governed by potential energy functions $U_1$ and $U_2$ is given by

\begin{equation}
F_1 = \int_0^1 \langle U_1 - U_2 \rangle_\lambda d \lambda + F_2
\end{equation}

where the integral is taken over a set of systems governed by a linear combination of the forces from the two systems, $U_\lambda = (1-\lambda)U_1 + \lambda U_2$. For numerical efficiency, the integration is taken in two steps: firstly from the full DFT system and to a system governed by a simple empirical potential chosen to closely resemble the dynamics of the DFT system, and then from the empirical system to the idealized noninteracting system. For our empirical potential we used a simple two-body spline-form potential generated by the force-matching methodology of Izvekov \emph{et al} \cite{izvekov} in combination with a harmonic potential anchoring oxygen atoms to their lattice sites. Appropriate volumes for the simulation cell were determined using the variable-cell methodology of Hernandez \cite{hernandez}.

\begin{figure}[htbl]
\includegraphics[width=7.5cm]{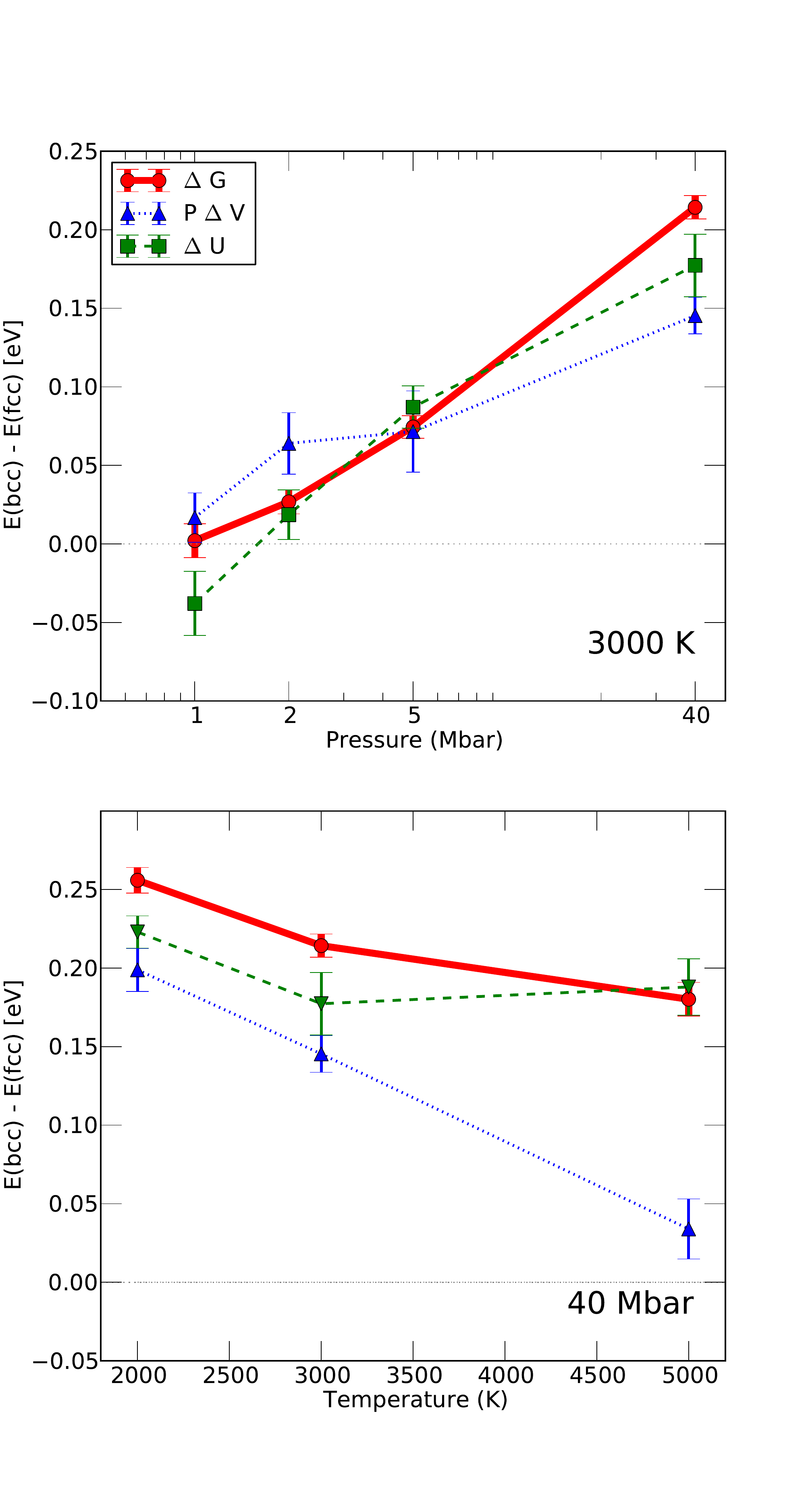}
\caption{Gibbs free energy difference per molecule between the bcc and fcc phases (solid line) shown as a function of pressure for a temperature of 3000~K, and as a function of temperature for a pressure of 40~Mbar. Also shown are the internal energy $\Delta U$ and pressure-volume $P \Delta V$ components of the free energy difference.}
\label{breakdown}
\end{figure}

\begin{table}
\begin{tabular} {c c | c c c c}
$P$(Mbar)& T(K) &  $\Delta G$ & $P \Delta V$ & $\Delta U $ & $-T \Delta S$ \\
& & [eV/mol] & [eV/mol] & [eV/mol] & [eV/mol] \\
\hline
1	&	3000	&	0.002(11)		&	0.017(16)		&	-0.037(20)	&	 0.023(28)		\\
2	&	3000	&	0.027(8)		&	0.064(20)		&	0.019(16)		&	-0.055(26)		\\
5	&	3000	&	0.074(7)		&	0.072(26)		&	0.087(13)		&	-0.084(30)		\\
10	&	5000	&	0.065(8)		&	0.054(24)		&	0.102(18)		&	-0.095(31)		\\
40	&	2000	&	0.256(8)		&	0.198(14)		&	0.229(10)		&	-0.165(19)		\\
40	&	3000	&	0.214(7)		&	0.145(12)		&	0.177(20)		&	-0.108(24)		\\
40	&	5000	&	0.180(11)		&	0.033(19)		&	0.188(18)		&	-0.042(28)		\\
\hline
\end{tabular}
\caption{Difference in free energy $G_{bcc} - G_{fcc}$ between the fcc and bcc phases, and the $P \Delta V$, $\Delta U$ and $T \Delta S$ components of the free energy difference.}
\label{free_energies}
\end{table}

Results of the free energy calculations are shown in Table \ref{free_energies}. We find the fcc structure to have a lower Gibbs free energy under all conditions studied, with the exception of the 1 Mbar and 3000~K point where the energy difference between bcc and fcc is nearly zero. Figure \ref{breakdown} plots free energy differences between the two structures as a function of pressure at 3000~K and of temperature at 40 Mbar. A strong tendency towards greater stability of the denser fcc structure as pressure is increased. We predict a transition from bcc to fcc stability at a pressure of $1.0 \pm 0.5$ Mbar. The error bar on the transition pressure means that the possibility that bcc may have no stability field at all is not excluded. 

The free energy difference between two structures may be broken down into three components; an internal energy term $\Delta U$, a volume term $P \Delta V$ and an entropic term $-T \Delta S$. The resulting components are listed in Table I. It is notable that the pressure-volume and internal-energy terms increasingly favor the fcc structure as pressure is increased, while the entropic term has the opposite sign.  

\begin{figure}[htbl]
\includegraphics[width=9.5cm]{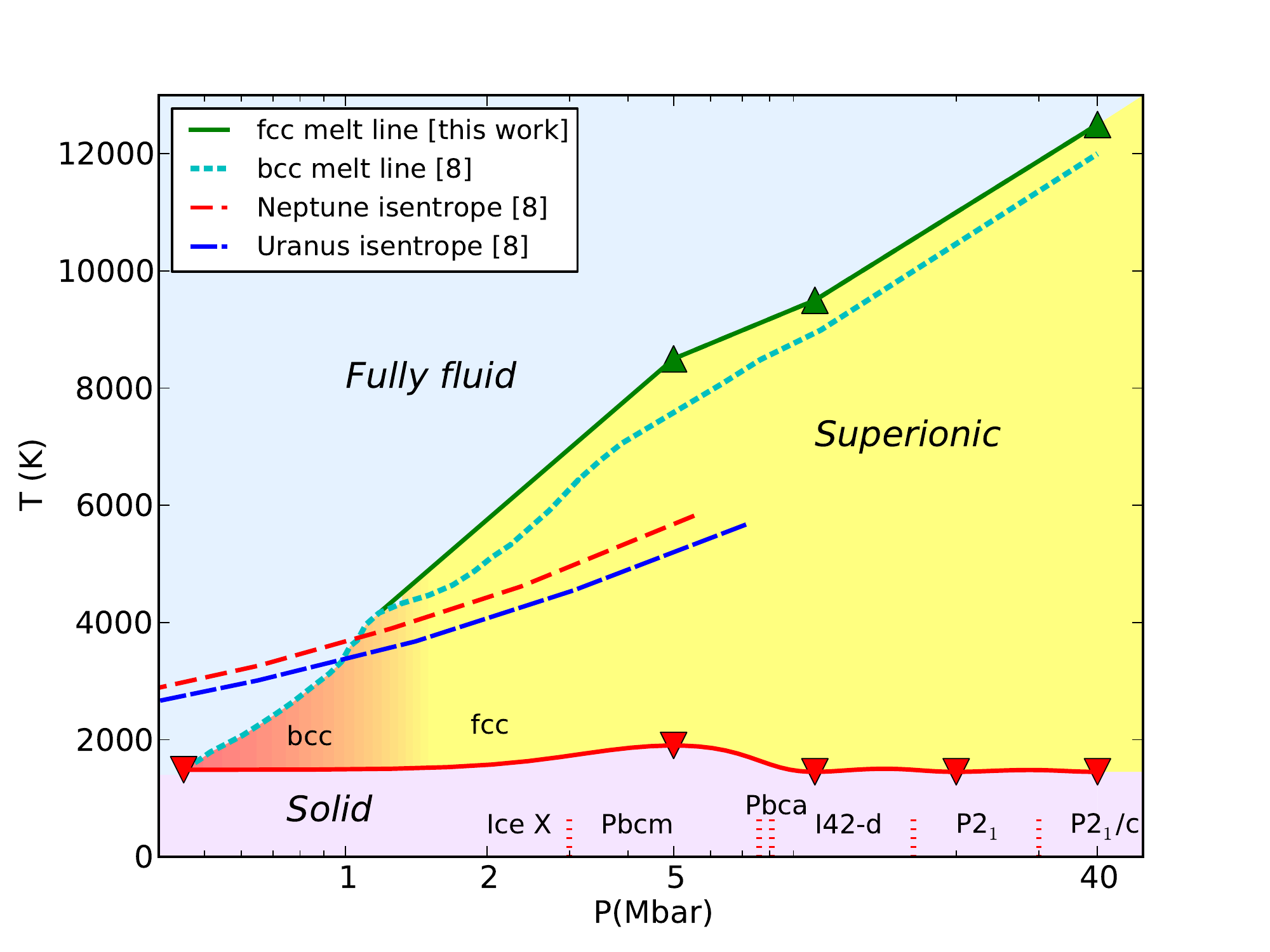}
\caption{Phase diagram of water showing the superionic regime. The phase boundary from solid ice to superionic water indicated by a shading gradient indicating the degree of uncertainty due to the error bars. The melting line of bcc superionic ice is from Redmer \emph{et al} \cite{redmer-icarus-11}, as are the indicated isentropes of Uranus and Neptune. The superionic regime is shaded to show the transition from bcc to fcc stability at pressures of $1 \pm 0.5$ Mbar. The fcc-superionic to fluid (red) and solid-to-superionic (green) lines were obtained by single-phase simulations in which the system was heated and cooled; we estimate a slightly higher melting temperature for the fcc lattice than was found by Redmer \emph{et al} for the bcc lattice.}
\label{phase}
\end{figure}

\begin{figure}[htbl]

\includegraphics[width=6.5cm]{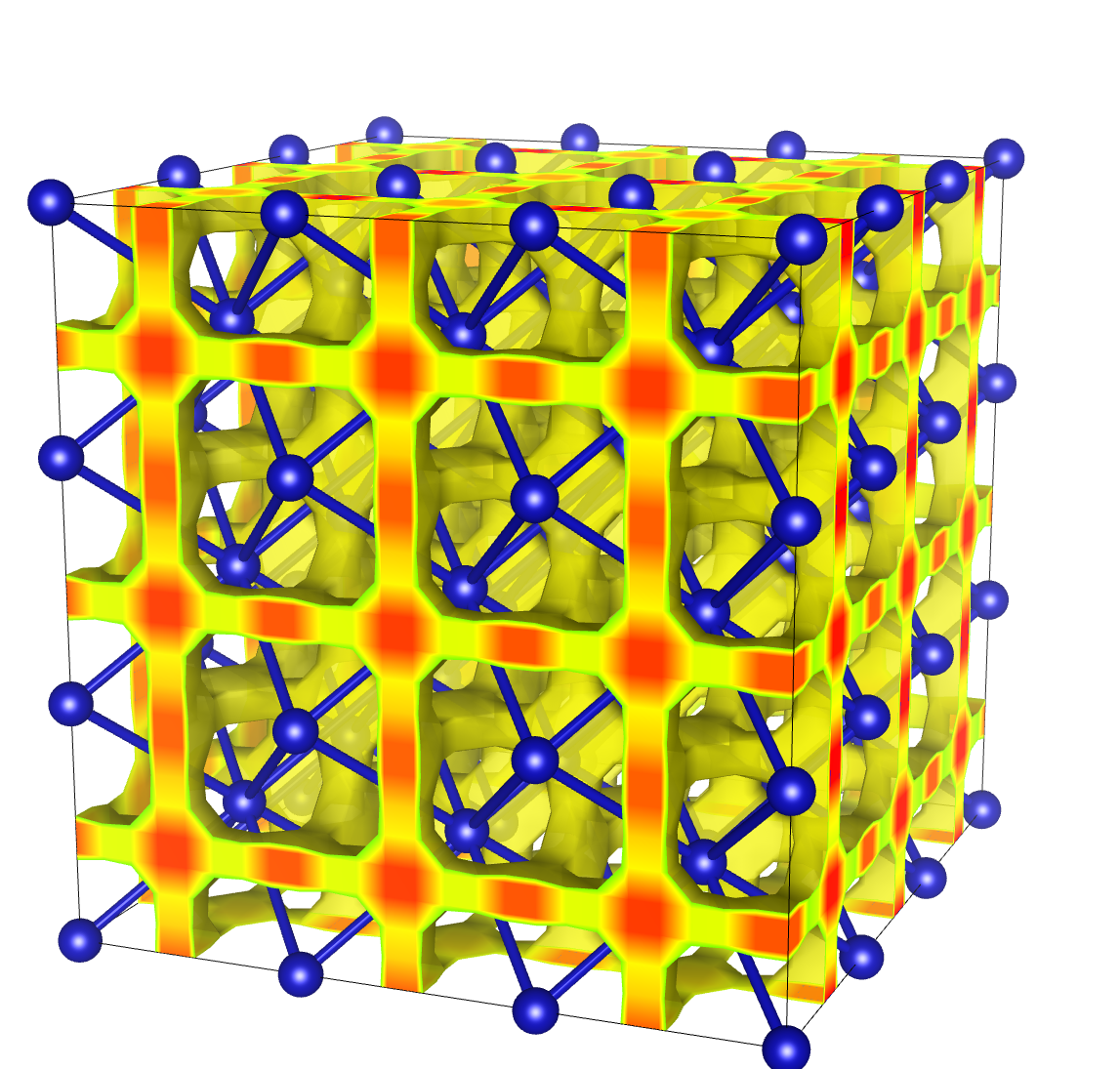}
\includegraphics[width=6.5cm]{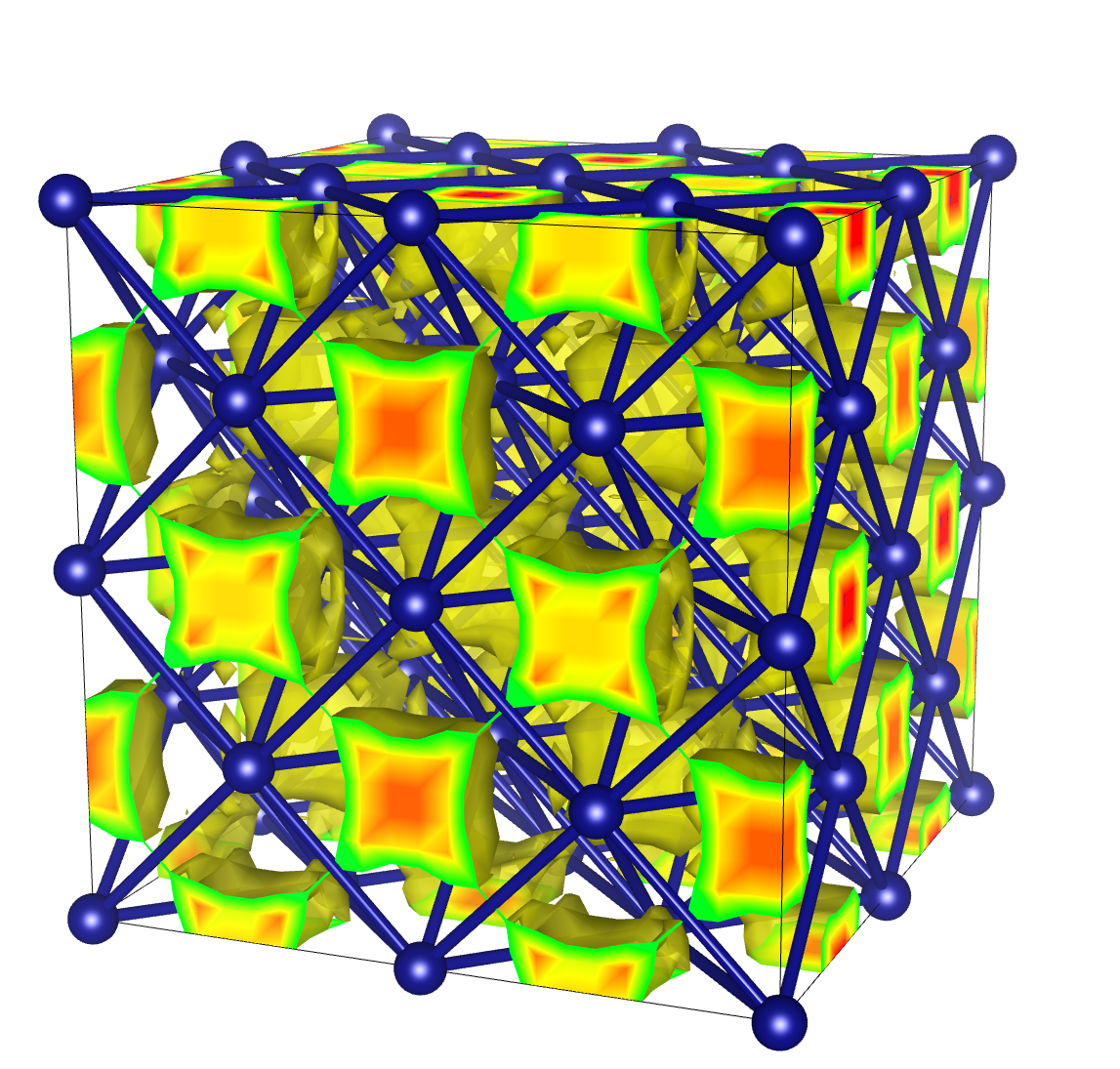}

\caption{Isosurfaces of constant hydrogen density in molecular dynamics simulations of superionic ice in the bcc (upper) and fcc (lower) phases at 40 Mbar and 5000~K. Two different maxima representing the tetrahedral and octohedral sites are seen in the bcc lattice; in the fcc lattice by contrast the hydrogens are concentrated at the tetrahedral interstitial sites. }
\label{isosurfaces}
\end{figure}

A key physical characteristic of the superionic phase is the hydrogen mobility, shown in Table II. We find hydrogen to diffuse more slowly in the fcc than the bcc structure under all conditions. The greater packing density of the fcc lattice allows fewer channels for hydrogens to diffuse from one site to another. Figure \ref{isosurfaces} shows isosurfaces of hydrogen density throughout molecular dynamics runs carried out at 40 Mbar and 5000~K. Apparent from these images is the fact that hydrogen atoms in the fcc structure are largely confined to tetrahedral intersititial sites, while in the bcc structure may migrate more freely between two different types of interstitial site (tetrahedral and octohedral). The greater variety of sites available to the hydrogen atoms in the bcc structure may also explain the entropic preference for the bcc structure. The transition from bcc to fcc thus coincides with a drop in hydrogen mobility, with consequences for thermal and electrical conductivity properties.

\begin{table}
\begin{tabular} {c c c c c}
$\rho$ & D(bcc) & D(fcc) & D(bcc) & D(fcc)\\
 (g/cc) & 2000K & 2000 K & 5000 K & 5000 K \\
 \hline
11	&	0.48	&	0.20	&	2.48	&	1.33	\\
9	&	0.79	&	0.48	&	3.81	&	2.18	\\
7	&	1.55	&	0.79	&	5.24	&	3.52	\\
5	&	1.20	&	0.98	&	6.49	&	4.23	\\
\hline								
\end{tabular}
\label{dtable}
\caption{Diffusion constants for hydrogen in bcc and fcc superionic ice, in \AA$^2$ fs$^{-1}$. }
\end{table}

We have established the stability of the fcc over the bcc superionic phase for pressures in excess of 1.0 $\pm$ 0.5 Mbar. As we have not considered all possible sublattice structures we cannot exclude the possibility of an alternative phase with lower Gibbs free energy across some or all of the phase diagram. Given the recent observation by X-ray diffraction of superionic water with a bcc sublattice by Sugimura \emph{et al} \cite{sugimura} at pressures up to 1 Mbar, we predict that a phase transition from the bcc phase to another superionic phase, whether fcc or some alternative structure, should occur within this pressure regime. Such a transition could be experimentally observed by laser heated diamond anvil cell \cite{goncharov-prl-05,sugimura} or laser-driven shock experiments \cite{lee-jchemphys-06}. The rearrangement of oxygen atoms can best be detected with X-ray diffraction or X-ray Absorption Near Edge Structure techniques. Figure \ref{phase} also implies that a superionic transition may occur along the isentropes of Uranus and Neptune, or that alternatively these planets may bypass the bcc superionic regime altogether. 

Previous theoretical studies of superionic water require some reconsideration in light of these results. The fcc phase has a higher melting temperature than the bcc phase, which will change computed equations of state for this material and may lead to a larger superionic ice regime in giant planet interiors than had previously been considered \cite{french-prb-09,redmer-icarus-11}. Interior models of Uranus and Neptune will require some revision, although the relatively small ($< 0.5 \%$) difference in density between the two phases may preclude a large effect.  The consequences of a superionic to superionic phase transition should be considered in the context of whether such a transition may help explain the non-axisymmetric non-dipolar magnetic fields of these two planets \cite{stanley}. The conclusions of our previous work on the solubility of water ice in metallic hydrogen within gas giant planet interiors \cite{wilson-astrop-12} do not change significantly, due to the relatively small magnitude of the free energy difference between superionic phases (~0.1 eV/molecule) compared to the large free energy (several eV/molecule) associated with solubility at Jupiter and Saturn core-mantle boundary conditions. Further work, including understanding the potential implications of phase changes in the superionic regime for the convective and heat transport properties of Uranus and Neptune, as well as experimental work aimed at detecting this phase change in practice, may provide further insight into the interiors of these poorly-understood ice giants.


\begin{thebibliography}{10}

\bibitem{militzer-prl-10}
B.~Militzer and H.~F. Wilson,
\newblock Phys. Rev. Lett. {\bf 105}, 195701 (2010).

\bibitem{mcmahon-prb-11}
J.~M. McMahon,
\newblock Phys. Rev. B {\bf 84}, 220104 (2011).

\bibitem{wang-nature-11}
Y.~Wang, H.~Liu, L.~Zho, and H.~Wang,
\newblock Nat. Comms.  (2011).

\bibitem{ji-prb-11}
M.~Ji, K.~Umemoto, C.-Z. Wang, K.-M. Ho, and R.~M. Wentzcovitch,
\newblock Phys. Rev. B {\bf 84}, 220105 (2011).

\bibitem{hermann-pnas-12}
A.~Hermann, N.~W. Ashcroft, and R.~Hoffmann,
\newblock Proc. Nat. Acad. Sci. {\bf 109}, 745 (2012).

\bibitem{french-prb-09}
M.~French, T.~R. Mattsson, N.~Nettelmann, and R.~Redmer,
\newblock Phys. Rev. B {\bf 79}, 054107 (2009).

\bibitem{redmer-icarus-11}
R.~Redmer, T.~R. Mattsson, N.~Nettelmann, and M.~French,
\newblock Icarus {\bf 211}, 798 (2011).

\bibitem{cavazzoni}
C.~Cavazzoni {\em et~al.},
\newblock Science {\bf 283}, 44 (1999).

\bibitem{goldman}
N.~Goldman, L.~E. Fried, I.-F.~W. Kuo, and C.~J. Mundy,
\newblock Phys. Rev. Lett. {\bf 1994}, 217801 (2005).

\bibitem{polian-prl-84}
A.~Polian and M.~Grimsditch,
\newblock Phys. Rev. Lett. {\bf 52}, 1312 (1984).

\bibitem{goncharov-prl-05}
A.~F. Goncharov {\em et~al.},
\newblock Phys. Rev. Lett. {\bf 94}, 125508 (2005).


\bibitem{wilson-astrop-12}
H.~F. Wilson and B.~Militzer,
\newblock Astrop. J. {\bf 745}, 54 (2012).

\bibitem{ninet-prl-12}
S.~Ninet, F.~Datchi, and A.~M. Saitta,
\newblock Phys. Rev. Lett. {\bf 108}, 165702 (2012).

\bibitem{caracas}
R. Caracas,
\newblock Phys. Rev. Lett. {\bf 101} 085502 (2008).

\bibitem{vasp}
G.~Kresse and J.~Furthm\"uller,
\newblock Phys. Rev. B {\bf 54}, 11169 (1996).

\bibitem{morales-pnas-09}
M.~A. Morales {\em et~al.},
\newblock PNAS {\bf 106}, 1324 (2009).

\bibitem{wilson-prl-10}
H.~F. Wilson and B.~Militzer,
\newblock Phys. Rev. Lett. {\bf 104}, 121101 (2010).

\bibitem{wilson-prl-12}
H.~F. Wilson and B.~Militzer,
\newblock Phys. Rev. Lett. {\bf 108}, 111101 (2012).

\bibitem{izvekov}
S.~Izvekov, M.~Parrinello, C.~J. Burnham, and G.~A. Voth,
\newblock J. Chem. Phys. {\bf 120}, 10896 (2004).

\bibitem{hernandez}
E.~R. Hern{\'{a}}ndez, A.~Rodriguez-Prieto, A.~Bergara, and D.~Alf{\`{e}},
\newblock Phys. Rev. Lett. {\bf 104}, 185701 (2010).

\bibitem{lee-jchemphys-06}
K.~K.~M. Lee {\em et~al.},
\newblock J. Chem. Phys. {\bf 125}, 014701 (2006).

\bibitem{stanley}
S.~Stanley and J.~Bloxham,
\newblock Icarus {\bf 184}, 556 (2006).

\bibitem{sugimura}
E. Sugimura {\em et~al.},
\newblock J. Chem. Phys. {\bf 137} 194505 (2012).

\bibitem{santra}
B. Santra {\em et~al.},
\newblock Phys. Rev. Lett. {\bf 107} 185701 (2011).

\end{thebibliography}

{\bf Acknowledgements:}
This work was supported by NASA and NSF.

\end{document}